\documentstyle[11pt]{article}

\title{Neutrino Cosmology and Limits on Extended Technicolor}
\author{Lawrence M. Krauss\thanks{Also Department of  Astronomy.
Address after July 1: Dept. of Physics, Case Western Reserve
University, 10900 Euclid Ave, Cleveland Ohio, 44106}, John
Terning\thanks{SSC fellow},
and Thomas
Appelquist\\ Sloane Physics Laboratory, Yale  University, New Haven,
CT 06511}
\begin{document}
\setlength{\baselineskip}{24pt}
\maketitle
\begin{picture}(0,0)(0,0)
\put(295,240){YCTP-P8-93}
\end{picture}
\vspace{-24pt}
\begin{abstract}
\setlength{\baselineskip}{18pt}
Using Big Bang Nucleosynthesis limits on the
number of light neutrinos we derive a cosmological lower bound of
$\approx 2$ TeV on the scale of
extended technicolor interactions (or any other new interactions) for
the
third family in models
where heavy gauge bosons couple to both left and
right-handed neutrinos.
\end{abstract}
\section{Introduction}
Recent work [1-4] on technicolor (TC)
and extended technicolor (ETC) theories
suggests that it may be possible to describe
the observed charged-particle mass spectrum while keeping flavor
changing neutral currents supressed, the $\rho$ parameter
close to 1, and $S$
small \cite{SandT}.  Of course, if ETC purports to be a
theory of mass, it must  also accomodate the fact that neutrinos are
light, or massless. Until  recently, perhaps because of the
formidable
problems associated with explaining the observed quark and charged
lepton masses, this latter challenge has received little attention.

The chief difficulty in such an effort is that it is
hard to construct ETC
models without right-handed neutrinos. While this may present a
challenge to
reproducing a realistic neutrino
mass spectrum, it also suggests that neutrino physics can place  new
cosmological
constraints on ETC models.  If new right-handed states are introduced
into the
theory, one must ensure that these states are not fully populated at
the time of Big Bang Nucleosynthesis (BBN), or else the predicted
helium abundance may exceed
observational limits.  This in turn places upper limits on the
reaction  rates
of right-handed neutrinos.  Since it is massive ETC gauge boson
exchange
that is responsible for the interactions of the right-handed
neutrinos, a new upper
limit on ETC interaction strengths
places a new lower bound on the scale where the ETC gauge group is
broken.
\section { Why Are Neutrinos Light?}
 The fact that only
extremely light left-handed neutrinos are seen
in nature is one of the most puzzling features of the quark-lepton
mass
spectrum. As we have noted, this poses special problems for
ETC theories. With right-handed neutrinos present in the theory,
there
is at least
one simple
explanation available for the fact that they have not yet been seen:
an
implementation of
the usual seesaw mechanism \cite{GMRS,models}.
The idea of the seesaw mechanism is that right-handed neutrinos
get large Majorana masses, so that the left-handed neutrinos end up
with masses
given by a Dirac mass squared divided by the Majorana mass.  It is
natural to assume that the
neutrino Dirac masses are of the same order as their charged leptonic
partners' masses. It is also natural to take the
Majorana masses to be of the same order as some ETC scale
(or scales) if condensates of bilinears of
right-handed neutrinos (Majorana condensates) are
involved
in the dynamical breaking of the ETC gauge symmetry. If there is
a hierarchy
of ETC scales to arrange
for a hierarchy of charged lepton masses ($m_e$, $m_\mu$, $m_\tau$),
each of
which is expected to be of order $(1  {\rm TeV})^3$ divided by the
square of the
corresponding ETC scale,
then the masses of the associated neutrinos are fixed at the same
time.
One finds that in this case
$ m_{\nu_{\mu}}  \approx O(10-50)  {\rm eV}~$, and $m_{\nu_{\tau}}
\approx O(10-50) {\rm keV}$ .
This mass spectrum is of course
incompatable with the MSW
solution of the solar neutrino problem.  More
importantly, $m_{\nu_{\tau}}$
is unacceptably large, requiring exotic decay mechanisms
in order to satisfy the constraints of cosmology \cite{Krauss}.
As a result, we will not consider this mechanism further in this
work.

A second possibility is that TC produces Dirac
 masses for the technineutrinos, but that these
masses do not feed down to the ordinary neutrinos.  One example
is a suggestion
made long ago by Sikivie, Susskind,
Voloshin, and Zakharov (SSVZ) \cite{Sikivie}.
SSVZ considered an $SU(3)_{ETC}$ model that describes the
interactions of one family of technifermions and the third family
of ordinary fermions. They concentrated only  on the lepton sector,
which contains the following
left-handed and charge-conjugated
right-handed fermions (where $E_R^c=(E_R)^c$)
labeled with their $SU(3)_{ETC}\otimes
SU(2)_L\otimes U(1)_Y$ charges:
\begin{eqnarray}
\left(\begin{array}{c}N_{L1} \\ E_{L1} \end{array}
\begin{array}{c}N_{L2} \\ E_{L2} \end{array}
\begin{array}{c} \nu_{\tau L} \\ \tau_L \end{array}\right)
\sim ({\bf 3},{\bf 2})_{-1}
& {\rm ;} &
N_{R1}^c, N_{R2}^c,\nu_{\tau R}^c  \sim
({\bf 3},{\bf 1})_{0}\nonumber \\
& & E_{R1}^c, E_{R1}^c, \tau_R^c  \sim
({\bf \overline{3}},{\bf 1})_{2}
\end{eqnarray}
down
It is straightforward to devise
mechanisms to break $SU(3)_{ETC}$ to $SU(2)_{TC}$, with the
addition of particles that also cancel $SU(3)_{ETC}$ anomalies
\footnote{The SSVZ model
suffers from an $SU(2)$ Witten anomaly \cite{Witten}, which can be
avoided by the addition of one multiplet.}.  Then when the TC
interactions  get strong, $E_L$ condenses with $E_R^c$, and  $N_L$
condenses with $N_{R}^c$.  Note that $N_L$ and $N_{R}^c$.
technifermions both transform as ${\bf 3}$'s under $SU(3)_{ETC}$, as
opposed to the $E_L$ and the $E_R^c$ which transform as a  ${\bf 3}$
and a ${\bf \overline{3}}$ respectively.  As a result, while
 the
standard one-ETC-gauge-boson-exchange graph
feeds down a mass to the $\tau$, the corresponding graph for the
$\nu_\tau$ vanishes for group theoretical reasons.  Thus the
$\nu_\tau$ remains massless.  Note that the
$\nu_{\tau R}^c $ in this model is sterile, aside from ETC
interactions.

This provides one example of the general class of models we
will constrain here, ie. those involving small (or vanishing) Dirac
masses for
the existing
neutrinos, with sterile right handed components save for new
interactions
mediated by heavy gauge bosons such as ETC gauge bosons.

\section{ETC, QCD, and BBN}
One might imagine that having light right-handed partners, Dirac or
Majorana, for
standard model neutrinos would destroy the agreement between
BBN calculations and observations of helium \cite{schramm},
which now suggest
that the number of equivalent light
neutrinos, $N_\nu$, is less than\footnote{In fact, if the
primordial helium abundance is constrained to be less than $24\%$, as
is sometimes argued, the limit is closer to 3.3 \cite{KraussRom}.  We
will conservatively assume here, however, that the
actual abundance can be somewhat larger, as the
existing error in this estimate is dominated by large systematic
effects.  Note that at present the BBN prediction is only consistent
if the
helium abundance is greater than $ \approx 23.6 \%$.}  about 3.5
\cite{KraussRom}.   However, extra neutrinos are
not  necessarily in thermal
equilibrium with electrons and photons at temperatures as low as 1
MeV. If the extra neutrinos have Dirac masses with the standard
left-handed $\nu$'s, then the combination of
weak interactions and helicity flips can contribute to thermal
population of right-handed states; one then finds that the Dirac mass
must be less than about 250 keV \cite{Krauss,Fuller}.  If the
neutrino
masses are smaller than this, population of the right-handed  states
can only occur by new interactions.  Since this proceeds by ETC gauge
 boson exchange, requiring that the right-handed states do not become
populated places
a lower bound on the mass of the ETC gauge boson divided by the ETC
gauge coupling.

If the decoupling temperature
of a particular neutrino is above the QCD phase transition
temperature ($T_{QCD} \approx 150$ MeV), then the subsequent
reheating of
 particles coupled to the radiation gas will cause such a decoupled
neutrino
to contribute  less than about 0.1 of a fully populated neutrino
state to the
radiation energy density below ($T_{QCD}$).  This is currently
compatible
with BBN limits.  We therefore investigate what ETC scale a
decoupling
temperature of $T_{QCD}$ corresponds to.  One can derive independent
bounds
in this manner for each ETC scale in  the  generic case where
separate scales
exist for each family.  However, in  such
models it is generally the third family that is associated with the
lowest
 ETC
breaking
scale.  We thus focus on the tau neutrino here.

There are
 two ETC processes which can populate right-handed tau neutrino
states
directly: $\nu_L \overline\nu_L \rightarrow \nu_R \overline\nu_R$ and
$\tau
\overline\tau \rightarrow \nu_R \overline\nu_R$. Of these
two processes only the former is significant at temperatures of order
$T_{QCD}$.  This is because the rate for these processes is
proportional to the
square of the density of incoming particles and at
$T \approx 150$ MeV, the number density of $\tau$ leptons is
suppressed
by
$\approx e^{-10}$ compared to neutrinos.  Thus, even though
the
cross section for $\tau$ annihilation at these temperatures is
$\approx 10^2$
times larger than for neutrino annihilation (due to the larger center
of mass
energies involved), their number density is
sufficiently small to make this reaction
insignificant.
The cross section for right-handed $\tau$ neutrino
production by ETC
interactions for center of mass energy $2E$ can then be
straightforwardly
derived to be:
\begin{equation}
\sigma = \left({g\over M}\right) ^4 {E^2\over 6 \pi} A^2~,
\label{eqn:cross}
\end{equation}
where $g$ is the gauge coupling, $M$ is the mass of the ETC gauge
boson, and $A^2$ represents a group theory
factor, which, for example, in the case of $SU(3)_{ETC}$ is
$1/9$.

To determine the rate of population of right-handed neutrinos at
temperature  $T$, one must solve a Boltzmann equation for their
number
density:
\begin{equation}
{d\eta_R \over dt}  = -3 {{\dot R} \over R} \eta + \langle \sigma v
 \rangle
\eta_L^2 - \langle \sigma v \rangle \eta_R^2 ~,
\label{eqn:boltz}
\end{equation}
where $\eta_L$ and $\eta_R$ are the number densities of left and
right-handed $\nu_\tau$'s, $R$ is the cosmological scale factor, and
the
angular brackets imply a thermal
average at temperature $T$.
The incident neutrinos are relativistic, so $v=1$, and it is
straightforward
to evaluate the thermal average of (\ref{eqn:cross}). One finds:
\begin{equation}
\langle \sigma v \rangle = {12.94 \,T^2 \over 6\pi} \left({g \over M}
\right)^4 A^2 ~.
\label{crossave}
\end{equation}

Next, following standard methods \cite{Weinberg} we can use
Einstein's equations for the Friedmann-Robertson-Walker expansion to
rewrite equation (\ref{eqn:boltz}) in a
form which is more easily solved.  Writing a
dimensionless quantity ${\tilde \eta} = \eta/T^3 $ one finds
\begin{equation}
{d{\tilde \eta_R} \over dT}  = -{\left({ 3\over 8 \pi G g*}
\right)^{1/2}}
\langle \sigma v \rangle ({\tilde \eta_L^2} -
 {\tilde \eta_R^2}) ~,
\label{eqn:bolttherm}
\end{equation}
where
\begin{equation}
g_* = {\pi^2 \over 30} \left(g_b +{7 \over 8} g_f \right)~,
\label{g*}
\end{equation}
gives the contribution of helicity states in the radiation gas to
the total energy density, and $g_b$ and $g_f$ are the number of boson
and
fermion helicity states in the radiation gas at temperature $T$.

 Now, define the thermal
equilibrium value,
\begin{equation}
{\tilde \eta}_{thermal} ={3 \over 4} {\zeta (3)
\over \pi^2}  =k,
\end{equation} for a single neutrino helicity state. After the QCD
phase transition at $ T_{QCD} \approx 150$
MeV, there is rapid  reheating of
all light or massless states in thermal equilibrium at that time.
Before any possible reheating of right-handed neutrinos (which we
are assuming are essentially decoupled) therefore,
it is reasonable to assume an initial condition $\tilde
\eta_R =\epsilon_i k$, where
\begin{equation}
\epsilon_i
={{ g_*(T<T_{QCD})}\over{ g_*(T>T_{QCD})}}
\approx 0.28 ~.
\end{equation}
 This takes into account the temperature
suppression of states not in thermal equilibrium during the period of
reheating following QCD the phase transition, which we assume is
adiabatic.  By
assumption the
right-handed neutrino states are not included in the helicity sums.
To
arrive at the value quoted, we include up, down, and strange quarks
in
our initial ($T>T_{QCD}$) radiation gas.

Given this initial condition,
we wish to see if the final abundance, $\tilde \eta_R$, rises to
a value
larger than its initial value after the QCD phase transition.  In
particular, we are interested in whether this  quantity  approaches
$\approx 0.5$ of a neutrino in thermal equilibrium, that is  ${\tilde
\eta}_R(T\approx1{\rm MeV})/k \approx 0.5$ (i.e. $ N_{\nu}  < 3.5$).
Using
equation
\ref{crossave}, we can  then integrate (\ref{eqn:bolttherm})
analytically.  Using our initial condition for $ \epsilon_i$, we find
\begin{equation}
\epsilon_f = {\tilde
\eta}_R(T_f)/k = \left[ {\left(1+\epsilon_i \over
1-\epsilon_i \right)} \exp \left( {2 \over 3} P [T_{QCD}^3 -
T_f^3] \right) -1 \over {{\left(1+\epsilon_i \over 1-\epsilon_i
\right)} \exp \left( {2 \over 3} P [T_{QCD}^3 - T_f^3] \right)
+1}\right]~,
\label{analytic}
\end{equation}
where
\begin{equation}
P= {12.94 k \over 6 \pi}
{\left({ 3\over 8 \pi G g*} \right)^{1/2}} {\left( g \over M
\right)}^4 A^2~.
\end{equation}

One can also derive a useful
approximate result, in which the dependence of $\epsilon_f$
on the fundamental parameters is clearer. This is obtained by
ignoring
the second term in  (\ref{eqn:bolttherm}).  In this case one finds:
\begin{equation}
\epsilon_f
 \approx \epsilon_i+{P\over 3} \left(T_{QCD}^3 - T_f^3\right) ~.
\label{simple}
\end{equation}
 This result will be good as
long as both $\epsilon_i$ and $\epsilon_f$ remain small.  We find
that
for $\epsilon_f <
1/2$, the bounds obtained using this result are in good agreement
with those
obtained
from (10).  Note that a larger $T_{QCD}$  would result
in a more stringent bound on the ETC scale, so $150$ MeV represents a
conservative choice.

In order to use these results to bound $M/g$ we must take two
further factors into account.  First, below the pion and muon
annihilation thresholds (when $ T \approx m_{\pi},m_{\mu}$)
right-handed
neutrinos can be further diluted.  We can account for this by
approximating the reheating due to pion and muon annihilations as
instantaneous and then integrating in stages.  We first integrate
to find $\epsilon_f$ just above the pion threshold, and then
calculate
the dilution of this quantity (as in eq. (8)) which would result from
reheating of the thermally coupled radiation gas due to pion
annihilations,
assuming  again that the
right-handed neutrinos are decoupled at this point.  We then
input this value as $\epsilon_i$ in a new integration (using either
(\ref{analytic}) or (\ref{simple})) down to $T=m_{\mu}$, using the
new appropriate $g_*$ value
in this range, and then repeat the above procedure, and
then integrate down to $ T\approx 1$ MeV.

Finally, we must remember that the quantity of interest is not
$\epsilon$ itself, but rather $\epsilon^{4/3}$.  This is
due to the fact that the quantity which is bounded by BBN is the
neutrino
energy density and not its number density.
The former is proportional to the $4/3$ power of the latter.

We display in figure 1 our final result for the quantity
$\epsilon_f^{4/3}$, based on succesive iterations of
(\ref{analytic}),
across the
pion/muon thresholds as described above.  If there is a single extra
right-handed neutrino, in order for it to
contribute less than $ \approx 0.5$ extra neutrino species to the
expansion rate during nucleosynthesis, then  $M/g > 2.2$~-~$4$ TeV,
depending on the value of $A^2$.  Note that where the
curve approaches unity the extra neutrino
is beginning to be in thermal equilibrium near $T_{QCD}$.

Thus, cosmology places a severe constraint on ETC
models of this type.   If we use the more restrictive BBN
bound of 3.3 effective neutrino species, then the lower bound on
$M/g$ would rise only slightly to $ 3$-$5$ TeV.  These results are
significant because ETC scales this small are likely to be required
to get a
$t$ quark mass above $100$ GeV.  Note that if a model
had 6 or more extra neutrino helicity states,
then at least some of the extra neutrinos would
have to decouple above the $\tau$ and $c$ quark thresholds (or have
rapid decays) in order
that $\epsilon_i < 0.1$. Using  equation (\ref{simple}), we find that
the lower  bound would then be pushed up to $M/g > O(20)$ TeV.

While our analysis was carried out in the context of one
class of models, the arguments are generic, and must be taken into
account in any ETC model which purports to explain neutrino masses,
or for
that matter in any model that has heavy gauge bosons which couple to
right
handed neutrinos. For example, the recent model of Randall
\cite{Randall} has $3$ extra right-handed neutrinos which form $3$
Dirac
particles with $3$ more sterile neutrinos.  The right-handed
neutrinos have
ETC interactions, but only with right-handed, up-type quarks.  Thus
below
$T_{QCD}$ these neutrinos will be produced by $\pi^+ \pi^-
\rightarrow \nu_R
{\overline \nu_R}$.  Below $T=m_\pi$, they are produced by $\gamma
\gamma
\rightarrow \nu_R{\overline \nu_R}$, and by bremsstrahlung off
protons and
neutrons (both reactions proceed through virtual $\pi^0$'s).  Since
the ETC
scale in this model is only $1.5$ TeV,  arguments of the type
presented here
could provide severe constraints on it, unless some mechanism allows
the
extra neutrinos to decay before the BBN era.

\vskip 0.15 truein
We would like to thank R. Sundrum and L. Randall for helpful
conversations.
This work was partially supported by the Texas National Laboratory
Research Commission, by the Natural Sciences and Engineering
Research Council of Canada, and by the Department of Energy under
contract \#DE-AC02ERU3075.
\vskip 0.15 truein


\vfill\eject

\noindent{Figure 1: Contribution of an extra neutrino to the energy
 density at the time of BBN expressed as a fraction of that due a
standard
model neutrino, as a function of the ETC scale. $A^2$ is a group
theoretic factor depending upon the ETC gauge group, and is $1/9$ for
$SU(3)_{ETC}$.
For comparison, the result is also displayed for a new U(1)
interaction,

where $A^2 =1$}

\end{document}